\documentclass[slac_one]{revtex4}

\usepackage{graphicx}
\usepackage{amsmath,amssymb}
\usepackage{fancyhdr}
\usepackage{psfrag}

\makeatletter       

\renewcommand{\p@subsection}{}

\makeatother

\newcommand{\LlR}{\ensuremath{{L\!/\!R}}}
\newcommand{\cve}{\ensuremath{c_V^e}}
\newcommand{\cae}{\ensuremath{c_A^e}}
\newcommand{\cvq}{\ensuremath{c_V^q}}
\newcommand{\caq}{\ensuremath{c_A^q}}
\newcommand{\Emiss}{\ensuremath{E\!\!\!\slash}}

\newcommand{\qw}{\ensuremath{\tilde{q}}}
\newcommand{\qwb}{\ensuremath{{\Tilde{q}^*}}}
\newcommand{\gw}{\ensuremath{\tilde{g}}}
\newcommand{\mqw}{\ensuremath{m_{\tilde{q}}}}
\newcommand{\mgw}{\ensuremath{m_{\tilde{g}}}}

\begin{document}

\author{A.~Brandenburg{\footnote{{\it Present address:} Genedata AG, CH-4016 Basel, Switzerland}}}
\affiliation{Deutsches Elektronen Synchrotron DESY, D-22603 Hamburg, Germany}
\author{M.~Maniatis}
\affiliation{Institut f\"ur Theoretische Physik, Philosophenweg 16, D-69120 Heidelberg, Germany }
\author{M.M.~Weber}
\affiliation{Department of Physics, University at Buffalo, Buffalo, NY14260-1500, USA }
\author{P.M.~Zerwas}
\affiliation{Inst. Theor. Phys. E, RWTH Aachen U., D-52074 Aachen, Germany}
\affiliation{Deutsches Elektronen Synchrotron DESY, D-22603 Hamburg, Germany }

{\phantom{Q}} \hfill {\tiny DESY 08-076}   \\
{\phantom{Q}} \hfill {\tiny HD-THEP-08-13}  \\
{\phantom{Q}} \hfill {\tiny PITHA 08/12}

\title{
Squarks and gluinos at a TeV $e^+e^-$ collider:           \\[2mm]
Testing the identity of Yukawa and gauge couplings in SUSY-QCD}

\begin{abstract}
Supersymmetry predicts the identity of Yukawa and gauge couplings
in the QCD sector: $q\tilde{q}\tilde{g} = \tilde{q}\tilde{q}g = qqg$. 
We examine whether the $q \qw \gw$~Yukawa coupling can 
be determined, by methods complementary to LHC, by analyzing squark-gluino 
final states at a TeV $e^+e^-$ collider.
\end{abstract}

\maketitle

%
\section{Introduction}
\label{sec-intro}

\noindent
While in QCD non-Abelian gauge invariance leads to equal couplings for
quarks and squarks to gluons, supersymmetry guarantees the identity of
the Yukawa coupling between squarks, quarks and gluinos with the gauge
coupling, cf. 
\cite{Fayet:1977yc, Dimopoulos:1981zb, Sakai:1981gr}:
\begin{equation}
\label{eq-1}
  {\hat{g}}_s (q{\tilde{q}}{\tilde{g}}) = g_s ({\tilde{q}}{\tilde{q}}g) = g_s (qqg)  \,. 
\end{equation}
This identity is preserved if supersymmetry is broken by soft terms,
{\it i.e.}~gaugino/scalar masses, and bi/tri-linear couplings between
scalar fields. The relation is crucial for the natural extension of the theory
from the electroweak scale to the Planck-scale without introducing
quadratic divergences, which are generated in the squark propagator, for instance, 
by the bosonic squark/gluon loops and which are canceled by the fermionic
quark/gluino loop. \\

Several methods can be exploited to test the identity (\ref{eq-1}) by measuring the
magnitude of the Yukawa coupling $q{\tilde{q}}{\tilde{g}}$. At the LHC
the production of squark pairs in quark-quark collisions which is mediated
solely by the interchange of gluinos, provides a classical instrument for
the measurement of the Yukawa coupling in SUSY-QCD~\cite{Freitas:2007fd}. In
practice, however, an ensemble of auxiliary measurements of decay
branching ratios is necessary, presumably requiring LC supplements, if
the coupling should be determined in an [almost] model-independent way.
\\

\begin{figure}[t] 
\centering
\includegraphics[width=0.4\linewidth, angle=270,clip]{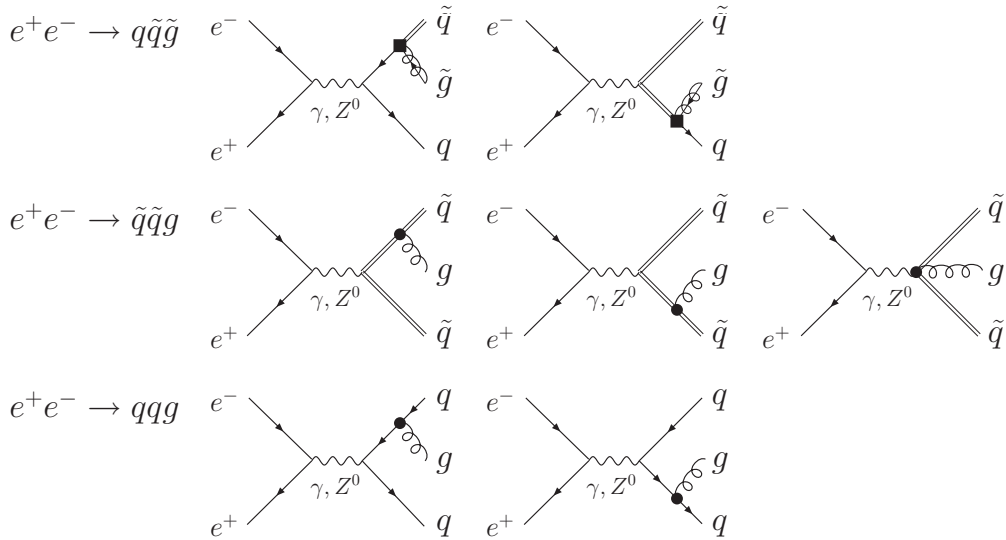}
\caption{\label{LOdiagrams}
{\it Generic Born diagrams corresponding to the processes~(\ref{eq-2a}) and (\ref{eq-2b}),(\ref{eq-2c})
which are proportional to the Yukawa coupling $\hat{g}_s$ [box] 
and to the gauge coupling $g_s$ [dot], respectively.}}
\end{figure}

A potential complement to this method is gluino emission in association with quark-squark
final states in $e^+e^-$ collisions, {\it cf.} Refs.~\cite{ILC,CLIC}:
\begin{align}
\label{eq-2a}
       e^+e^- &\to q \qw \gw\\
\intertext{which requires a non-zero Yukawa coupling. 
This process is related by
supersymmetry directly to the gauge processes of gluon radiation off squarks,}
\label{eq-2b}
      e^+e^- &\to {\tilde{q}} {\tilde{q}} g \,,\\
\intertext{and standard gluon radiation off quarks,}
\label{eq-2c}
      e^+e^- &\to q q g \,.                                   
\end{align}
Generic diagrams are displayed in 
Fig.~\ref{LOdiagrams}.
[Note that the abbreviations $qq \, ..$, used for notational clarity, should
anywhere be interpreted as the incoherent sum of particle-antiparticle
plus antiparticle-particle states of all flavors and $L,R$ indices.] It can be anticipated
without analyzing details that the process~(\ref{eq-2a}) will be useful in practice
only if the squarks and gluinos are moderately light; the range, however, is
quite compatible with electroweak precision analyses, cf. Ref.~\cite{ellisea}. 
The process is suppressed by the [small] Yukawa coupling squared, the propagator
transporting high (virtual) masses, and the
3-body phase-space involving two heavy particles; the suppression
remains effective even significantly
above the threshold.{\footnote{These difficulties contrast
with measurements of the Yukawa
couplings in the electroweak SU(2)$\times$U(1) sector in which chargino,
neutralino and selectron pair-production in polarized $e^+e^-$ and
$e^-e^-$ processes allow the determination of the electron-related Yukawa
couplings at the per-cent down to the per-mill level~\cite{Freitas:2003yp}.}}
Nonetheless, the production cross sections are in general
expected still to be considerably larger than loop-mediated gluino-pair
production in $e^+e^-$ collisions \cite{Berge:2002ev}. \\

Since any method for measuring the SUSY-QCD Yukawa coupling
is connected with ancillary problems, it is worth studying to what extent
the process~(\ref{eq-2a}) could be exploited to approach 
a solution of this experimental challenge. Since the
identity of Yukawa and gauge couplings is a central concept of
supersymmetric theories, it is mandatory to study all opportunities which
could shed light on this aspect. To reach solid conclusions, the theoretical analysis
has been performed including SUSY-QCD corrections at next-to-leading order [NLO] which
increase the complexity of the theoretical work enormously. 
In turn, this allows us to study the sensitivity of SUSY-QCD 
corrections in $\tilde{q} \tilde{q}$, $qq$ final states 
to the Yukawa coupling. 
Preliminary results had been contributed to conference proceedings 
\cite{Brandenburg:2002ff}. \\

%
\section{Squarks and gluinos in $\mathbf{e^+e^-}$ collisions}
%
\label{sec-2}

\noindent
Before discussing the dependence of individual processes on the SUSY-QCD
Yukawa coupling, an overview should be given on the size of cross
sections which can naturally be expected for experimental analyses. In
addition, the key points of the crucial techniques used in the higher-order
calculations are summarized. For the reader's convenience, we include
some earlier material from the literature as to provide a coherent
presentation of the problems. \\

The topology of the final states after the decays of the supersymmetric
particles down to the LSP [assumed to be the lightest neutralino
${\tilde{\chi}}^0_1$ in the present R-parity conserving set-up] depends
strongly on whether squarks are lighter or heavier than gluinos. In the
first scenario, squarks decay preferentially to charginos/neutralinos,
while gluinos decay to squark-quark pairs, followed by the subsequent
squark decays,
\begin{alignat}{1}
\begin{split}
\mqw < \mgw :\quad 
& \qw \to q + \tilde\chi        \to j  + E\!\!\!\slash \text{, etc}\\
& \gw \to \qw + q \to q + q + \tilde\chi      \to j j + E\!\!\!\slash \text{, etc}  \,.
\end{split}
\intertext{In the second scenario squarks decay to gluinos which subsequently
decay, mediated by virtual squarks, to charginos/neutralinos,}
\begin{split}
\mqw > \mgw :\quad
& \qw \to \gw + q \to \qw_{\text{virt}} + q + q \to q + q + q + \tilde\chi
               \to j j j + E\!\!\!\slash \text{, etc}\\
& \gw \to \qw_{\text{virt}} + q \to q + q + \tilde\chi \to j j  + E\!\!\!\slash \text{, etc} \,.
\end{split}
\end{alignat}
\begin{table*}[t]
\centering
\[\begin{array}{cr@{\;}c@{\;}lcccc}
\hline\hline
   &  \multicolumn{3}{c}{\text{Energy window}} 
    & \multicolumn{4}{c}{\text{Final states with strong gauge / Yukawa couplings}} \\
  & & & & & \alpha_s & \hat{\alpha}_s & \hat{\alpha}_s^2\\ 
  \hline
  &  2 \mqw \leq &\sqrt{s}& \leq \mqw+\mgw
    & \qw \qw (2j + \Emiss)&   \qw \qw g (3j + \Emiss) &  \\
  \mqw < \mgw \quad & \mqw+\mgw \leq &\sqrt{s}& \leq 2 \mgw
     & \qw \qw  (2j + \Emiss)&   \qw \qw g  (3j + \Emiss) &   q\qw\gw (4j + \Emiss) \\
  &  2 \mgw \leq &\sqrt{s}&  
    & \qw \qw (2j + \Emiss) &   \qw \qw g (3j + \Emiss)  & q\qw\gw (4j + \Emiss) & q q \gw\gw (6j + \Emiss) \\ \hline
  &  2 \mgw \leq &\sqrt{s}& \leq \mqw+\mgw
    & & & & q q \gw \gw (6j + \Emiss)  \\
  \mqw > \mgw \quad &  \mqw+\mgw \leq &\sqrt{s}& \leq 2 \mqw
    &  &  & {q\qw\gw}  (6j + \Emiss) & \langle q q \gw \gw  (6j + \Emiss) \rangle  \\ 
  &   2 \mqw \leq &\sqrt{s}&  
    &  \qw\qw (6j + \Emiss) &  \qw \qw g  (7j + \Emiss) & \langle q \qw \gw  (6j + \Emiss) \rangle
                                                        & \langle q  q  \gw \gw  (6j + \Emiss) \rangle \\ 
  \hline\hline
\end{array}\]
\caption{\label{tab:proc-regions}
  {\it Kinematically accessible processes involving squarks and gluinos
  at different c.m. energies 
  for $\mqw < \mgw$ and $\mqw > \mgw$. Processes
  are identified by the jet topology of the final states. The corresponding gauge and Yukawa
  couplings are denoted by $\alpha_s = g^2_s/4\pi$ and
  $\hat{\alpha}_s = {\hat{g}}_s^2/4\pi$, respectively. [Final states in $\langle ... \rangle$ 
  brackets in the second part of the table are generated primarily by squark decays
  to gluinos to which the power counting of the couplings does not apply.]}}
\end{table*}

In contrast to QCD jets, the jets in heavy squark/gluino decays are
well separated in phase space, and clustering will occur only at random.
A large fraction of the charginos/neutralinos $\tilde\chi$ is in general light 
and, if not the LSP, they decay to leptons [with taus generating only slim jets 
with up to three hadrons which we keep separate in the jet counting.] \\

A survey of all squark and gluino production processes relevant
for investigating the strong gauge and Yukawa couplings
is given in Tab.~\ref{tab:proc-regions}
which indicates the threshold energies and the jet topologies
for the individual continuum processes in which $\tilde{q}$ or $\tilde{g}$ are not
mutual decay products [QCD coupling $\alpha_s = g^2_s/4\pi$; Yukawa coupling
${\hat{\alpha}}_s = {\hat{g}}_s^2/4\pi$].
At the time when the measurement of the Yukawa coupling can be performed,
the masses of the supersymmetric particles will be determined very
precisely, {\it cf.} Ref.~\cite{lhclc}, and it will be evident which part of the
table is relevant. \\

It may be noted that final states of events generated by the decay of on-shell squarks
to quarks and the lightest neutralino as LSP, $\qw \to q + {\tilde{\chi}}^0_1$, can be 
reconstructed up to a two-fold ambiguity, an important identification tool
when the squark and neutralino masses are known, {\it c.f.} Ref.~\cite{hagi}. 
As the $\qw\qw$ c.m. frame [denoted by $\ast$] coincides with the laboratory frame, or,
can be reconstructed after gluon emission, the $\qw\qw$ axis in that frame is given 
by one of the two intersections of the cones centered at the quark axes with opening angles 
determined from $m^2_{\tilde{\chi}} = m^2_{\tilde{q}} - 2 E^\ast_q E^\ast_{\tilde{q}} 
(1-\beta^\ast_{\tilde{q}}\cos\theta^\ast_{\tilde{q}})$. The false solutions 
for the axes give rise to essentially flat background angular-correlations. \\          

\begin{figure}[h]
\centering
\begin{minipage}[b]{0.49\linewidth}
\includegraphics[width=\linewidth]{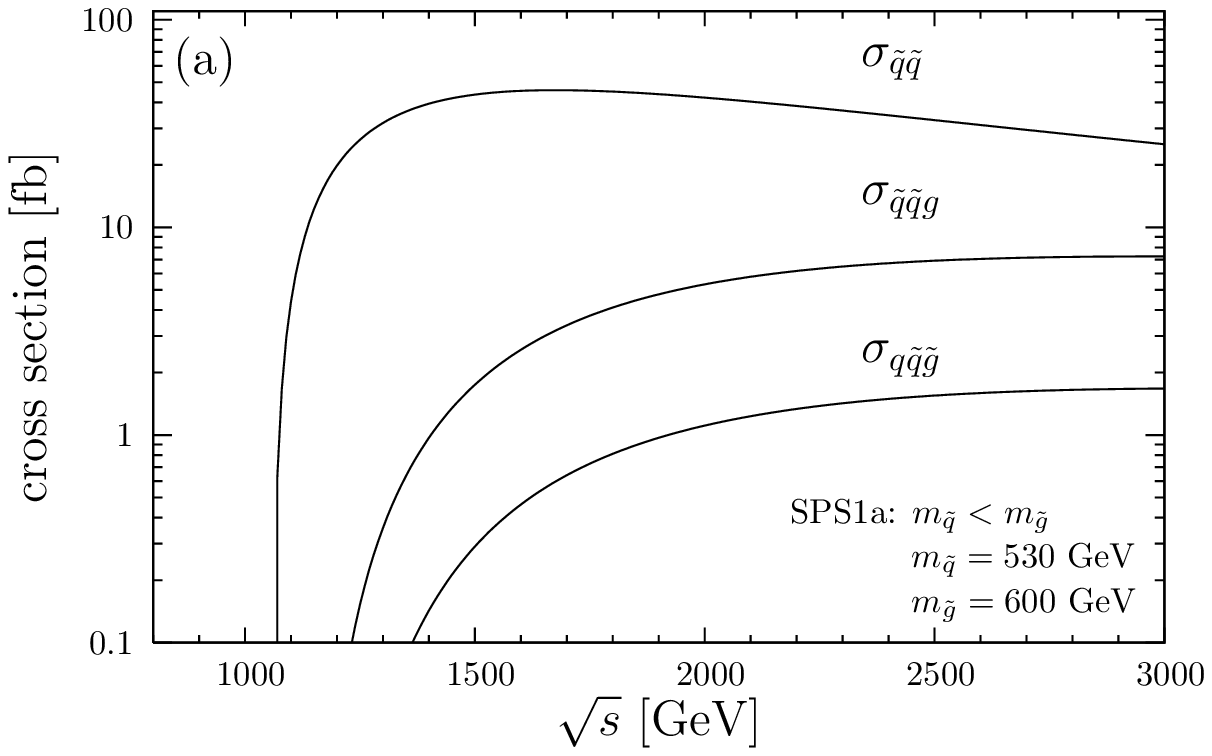}
\end{minipage}
\hfill
\begin{minipage}[b]{0.49\linewidth}
\includegraphics[width=\linewidth]{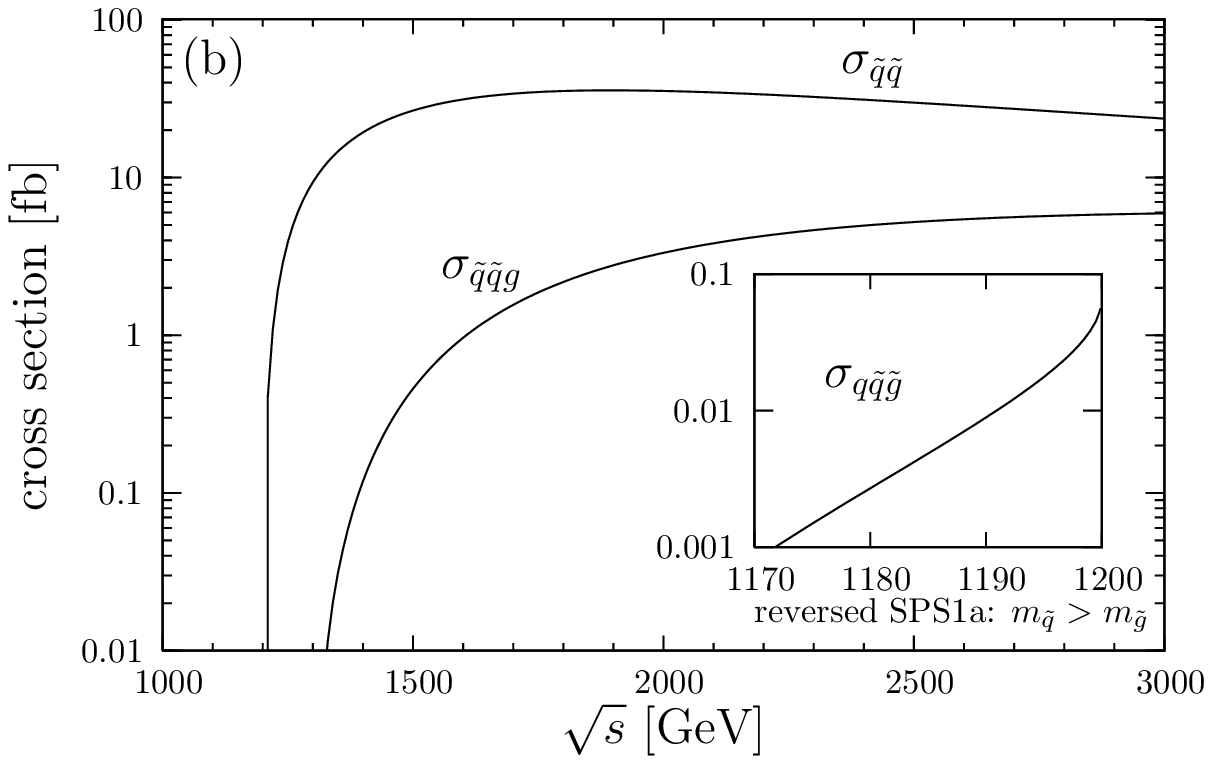}
\end{minipage}
\caption{\label{fig:LOcross} {\it Relevant tree level cross sections for the production
of squarks and gluinos. In Fig.(a) the gluino is assumed to be heavier than the
squarks, according to the SPS1a scenario, and in Fig.(b) the {\em reversed} mass scenario
is considered, where the gluino is lighter ($\mgw=530$~{\rm{GeV}}) than the squarks
(degenerate squark masses $\mqw=600$~{\rm{GeV}}).}}
\end{figure}
Leaving out rare processes of second order in the Yukawa coupling, the
cross sections for squarks lighter than the gluino are displayed in
Fig.~\ref{fig:LOcross}(a), summed over five flavors.
For clear identification a minimal energy cut \mbox{$E_X >$ 100 GeV} 
is applied to the gluon jet (details are given in the following subsections).
In Fig.~\ref{fig:LOcross}(a)
production cross sections are shown for
squark and gluino masses adopted from the standard SPS1a scenario~\cite{Snowmass},
i.e. $\mgw=600$~GeV and a common squark mass of $\mqw=530$~GeV,
[which is compatible with the mass limits allowed by the
searches for supersymmetric particles at the Tevatron~\cite{Affolder:2001tc}].
The dominant cross section of the $\qw \qw$ final state rises steeply
above the threshold proportional to the
third power of the velocity of the squarks. 
Since the gluino is heavier than all squarks in the SPS1a scenario, 
the production threshold for the $q \qw \gw$ final state is higher
than the $\qw \qw$ threshold. 
The jet topologies
in the final states are characteristic for the individual processes and
jet counting can be used as a powerful discriminant. [Of course, this 
Born approach must be refined by properly controlling QCD
showers in experimental analyses.] \\

The relevant cross sections for the scenario in which the squarks are
heavier than the gluinos, are shown in Fig.~\ref{fig:LOcross}(b). 
In a merely {\it ad hoc}
procedure, as any details beyond SUSY-QCD are irrelevant for the present
discussion, the mass values of squarks and gluino in SPS1a are just
reversed. [Such a scenario is actually close to SPS4.] 
Note that above the c.m. energy of twice the squark mass the
process $q \qw \gw$ proceeds primarily through squark pair production 
with subsequent decay of a squark into a gluino. 
The energy window for $q \qw \gw$ production {\it sui generis} 
is narrow and the cross section of the process, due to phase space suppression 
near the threshold, is rather small. \\

However, an additional opportunity for measuring the Yukawa coupling in
the second scenario is provided by the squark decays 
to gluinos, Ref.~\cite{Beenakker:1996dw}. 
The threshold behavior of the squark-excitation cross section in $e^+e^-$
collisions depends sensitively on the total decay width
$\Gamma(\tilde{q})_{tot}$. This method of measuring the total width
was discussed in Ref.~\cite{Freitas:2003yp} for slepton production but it can be
transferred to squarks in the same way; in fact, the impact of the total
width on the production cross section for squarks is significantly bigger as a
result of the increased size of the width. On the other hand,
the electroweak continuum cross section for squark-pair production with
squarks decaying to gluinos is proportional to the branching ratio
$BR(\tilde{q} \to q \tilde{g})$ squared. By combining the two threshold
and continuum measurements of total width and branching ratio, respectively,
\begin{equation}
    \Gamma (\qw \to q \gw) = BR(\qw \to q \gw) \times \Gamma(\qw)_{\text tot}
                       = \frac{\hat{g}^2_s}{6 \pi} \mqw \left(
		       1-\frac{\mgw^2}{\mqw^2} \right)^2                 \,,
\end{equation}
the SUSY-QCD Yukawa coupling can be determined from the partial width. \\[3mm]

Since in the analysis of the individual channels higher-order SUSY-QCD
corrections will be included, the technical set-up should be summarized
globally. To take advantage of our extensive infra-structure, the
dimensional regularization has been carried out in the DREG scheme. To
comply with the SUSY-QCD Ward identities, which are violated by the
mismatch of fermionic and bosonic degrees of freedom in this scheme,
counter terms must be added which however can be mapped onto a
modified relation between gauge and Yukawa couplings~\cite{Beenakker:1996dw,
Hollik:2001cz}:
\begin{equation}
\label{eq:Yukawa-counter}
  \hat{g}_s = g_s \left[ 1 + \frac{\alpha_s}{4\pi} 
    \left( \frac{2}{3} C_A - \frac{1}{2} C_F \right) \right]
  = g_s \left[ 1 + \frac{\alpha_s}{3\pi} \right]\,.
\end{equation}
The masses of the heavy particles, squarks and gluinos, are introduced as
pole masses. For the renormalization of the couplings the $\overline{\rm
MS}$ scheme is adopted. The ``experimental'' QCD gauge coupling
$\alpha_s^{(5)}(Q^2)$ is introduced for five massless quarks and
all heavy particles are decoupled. From this coupling, conventionally fixed
at the energy $Q = M_Z$, Ref.~\cite{Bethke}, the
appropriate QCD gauge coupling for the SUSY system, $\alpha_s(Q^2)$, above the
heavy particle thresholds can be derived as
\begin{align}
  \alpha_s(Q^2)= \alpha_s^{(5)}(Q^2)
  \bigg[ 1+ \frac{\alpha_s^{(5)}(Q^2)}{2 \pi}
            \left(  \frac{1}{3} \ln \frac{m_t^2}{Q^2}
                  + \frac{N_f}{6} \ln \frac{m_{\tilde{q}}^2}{Q^2}
                  + \frac{N_C}{3} \ln \frac{m_{\tilde{g}}^2}{Q^2}
            \right)
  \bigg]^{-1}
\end{align}
with
the number of flavors $N_f=6$ and colors $N_C=3$.
This procedure guarantees that the coupling
$\alpha_s(Q^2)$ smoothly approaches the asymptotic form
when $Q^2 \gg M^2$ for all masses $M$ of the
supersymmetric theory. In the calculations presented here 
we identify the renormalization scale with the c.m. energy.\\

After this introductory discussion of the individual channels 
and their dependence on the
SUSY-QCD Yukawa coupling, detailed evaluations are presented 
in the following subsections. \\

%
\subsection{Final-state supersymmetric particles}
\label{sec21}

%
\subsubsection{Squark pairs $\mathbf{\qw \qw}$}
%
\label{sec211}

\noindent
It has been shown in Figs.~\ref{fig:LOcross} that the cross section for squark-pair
production
\begin{equation}
   e^+e^- \to \tilde{q} {\tilde{q}}^\ast
\end{equation}
is maximal in the group of channels suited for the measurement of the SUSY-QCD
Yukawa coupling. The cross section, in Born approximation,
\begin{equation}
\sigma_{LO}(e^+e^- \rightarrow \qw_{\LlR} \; \qw_{\LlR}^\ast)= 
\frac{\pi \alpha^2}{s} C_{\LlR} \;\beta^3,
\end{equation}
rises as the third power of the $\tilde{q}$ velocity $\beta=(1-4\mqw^2/s)^{1/2}$ above
the threshold as demanded for $P$-wave production \cite{hagi}. 
The $L/R$ couplings,
\begin{equation}
\label{eq:C1-def}
  C_{\LlR} = Q_q^2 + ({\cve}^2+{\cae}^2) (\cvq\pm\caq)^2 \chi_Z^2 
    - 2 Q_q \cve (\cvq\pm\caq) \chi_Z
\end{equation}
are given by the electric and isospin charges of the squarks,
$c_V^f = (I_3^f - 2 Q_f s_W^2)/s_{2W}$ and
$c_A^f = I_3^f/s_{2W}$, 
where $s_W^2= {\sin}^2\theta_W$ is the electroweak mixing parameter, 
$s_{2W}=\sin{2{\theta}_W}$, and $\chi_Z = s/(s-m_Z^2)$ the scaled
$Z$-propagator. \\

\begin{figure}[h!t] 
\centering
\includegraphics[width=0.6\linewidth]{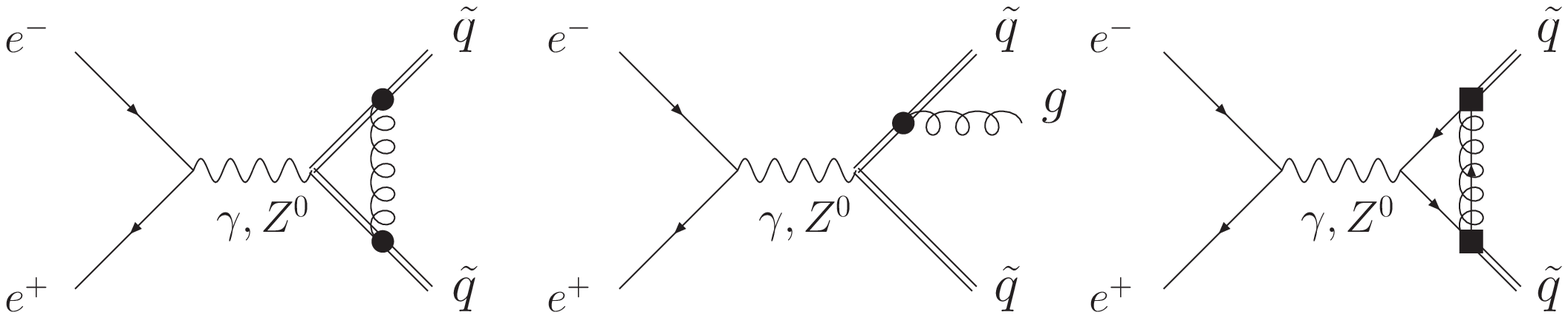}
\caption{\label{diag-sqsqDelta} {\it
Examples for diagrams of SUSY-QCD corrections to
the production of squark pairs as defined in (\ref{eq-twosusy}); left:
gluon vertex correction, mid: gluon bremsstrahlung;
right: gluino vertex correction. }}
\end{figure}

\begin{figure}[h!t] 
\centering
\includegraphics[width=0.45\linewidth]{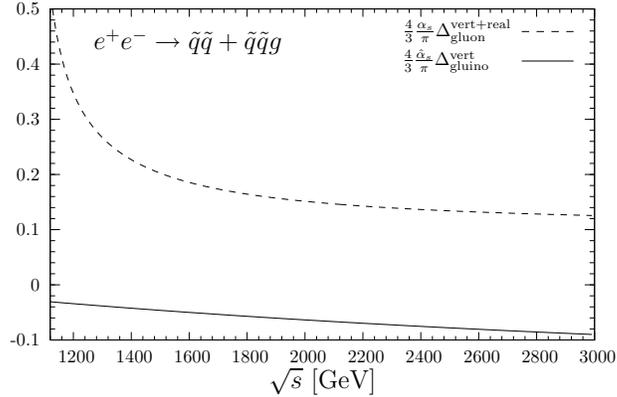}
\caption{\label{fig-sqrts-sqsqDelta} {\it
The NLO SUSY-QCD ($\alpha_s$ and ${\hat{\alpha}}_s$) corrections for squark pair production.
Separately shown are the sum of virtual and real gluon corrections, 
$(4 \alpha_s / 3 \pi) \Delta^\text{vert+real}_\text{gluon}$, as well as the gluino vertex correction,
$(4 {\hat{\alpha}}_s / 3 \pi) \Delta_\text{gluino}^\text{vert}$. Parameters correspond to the reference 
point SPS1a.}}
\end{figure}

The production cross section for squark pairs in next-to-leading order (NLO) includes
the gluon vertex correction and gluon bremsstrahlung, as well as the gluino vertex
correction, as depicted in Fig.~\ref{diag-sqsqDelta}. [Final states $q \qw \gw$ are
assumed to be separated explicitly.] The size of
these corrections,
\begin{equation}
\label{eq-twosusy}
  \sigma_\text{NLO}(e^+e^- \to \qw \qw^\ast + \qw \qw^\ast g) =
  \left.\sigma(e^+e^- \to \qw \qw^\ast)\right|_\text{Born} 
  \left[1+ \frac{4}{3} \frac{\alpha_s}{\pi} \Delta^\text{vert+real}_\text{gluon} + 
           \frac{4}{3} \frac{{\hat{\alpha}}_s}{\pi} \Delta^\text{vert}_\text{gluino}   \right] \,,
\end{equation}
depends on the masses of the squarks and gluinos, {\it cf.}
Ref.~\cite{Arhrib:1994rr}. 
The two parts contributing to the corrections are displayed,
for the set of SPS1a parameters defined earlier, in Fig.~\ref{fig-sqrts-sqsqDelta}.
The gluon vertex contribution is formally divergent at the threshold, {\it i.e.}~linearly
in the inverse velocity of the squarks, but this Coulomb
singularity is regularized by the non-zero squark width \cite{FreiFad}.
[{\it In toto}, the Born threshold suppression $\sim \beta^3$ is reduced to 
$\sim \beta^2$.] 
Compared to the Born cross section, ${\mathcal{O}}(50\,{\rm{fb}})$, the corrections,
in particular the gluino corrections involving the Yukawa coupling, are of order
femtobarn. With an integrated luminosity $\sim$ 1 ab$^{-1}$, these corrections
can clearly manifest themselves beyond statistical fluctuations. \\

%
\subsubsection{Squark-pair plus gluon final state $\mathbf{\qw \qw g}$}
%

\label{sec-212}

\noindent
A subset of final states in~\ref{sec211} is generated by the process
\begin{equation}
   e^+e^- \to \qw \qw^\ast g
\end{equation}
in which the gluon jet is isolated explicitly. This process measures the
QCD coupling of gluons to squarks. The differential cross section in the
scaled Dalitz variables $x_{\qw} = 2 E_{\qw} / \sqrt{s}$, etc,  
was calculated already some time ago \cite{Dahmen:1983xe}:
\begin{equation}
\label{LOsqsqgl}
  \frac{d\sigma ( e^+e^- \rightarrow \qw_{L/R} \; \qw_{\LlR}^\ast\; g)}
  {dx_{\qw}\:dx_{\qw^\ast}} = 
    \frac{\alpha^2 \alpha_s}{4s} C_{\LlR} 
   \biggl[
   16 - \frac{16}{3}\frac{\mu_{\qw}^2  y_{\qw}^2}{(1-x_{\qw})^2}
   -\frac{16}{3}\frac{\mu_{\qw^\ast}^2  y_{\qw^\ast}^2}{(1-x_{\qw^\ast})^2}
   -\frac{8}{3} \frac{(2 \mu_{\qw}^2 -1)(y_{\qw}^2+y_{\qw^\ast}^2-y_{g}^2)}{(1-x_{\qw})(1-x_{\qw^\ast})}
     \biggr]\,.
\end{equation}
To simplify the notation, the abbreviations $y_{\tilde{q}} = 2 |{\mathbf p}_{\tilde{q}}|/ \sqrt{s},
\; \mu_{\tilde{q}} = m_{\tilde{q}}/\sqrt{s}$, etc, have been introduced.
If the squarks are lighter than the gluinos, as realized in SPS1a, the
final state consists of three well-separated jets and a pair of neutralinos/charginos.  \\

\begin{figure}[h!t] 
\centering
\vspace{5mm}
\includegraphics[width=0.65\linewidth,clip]{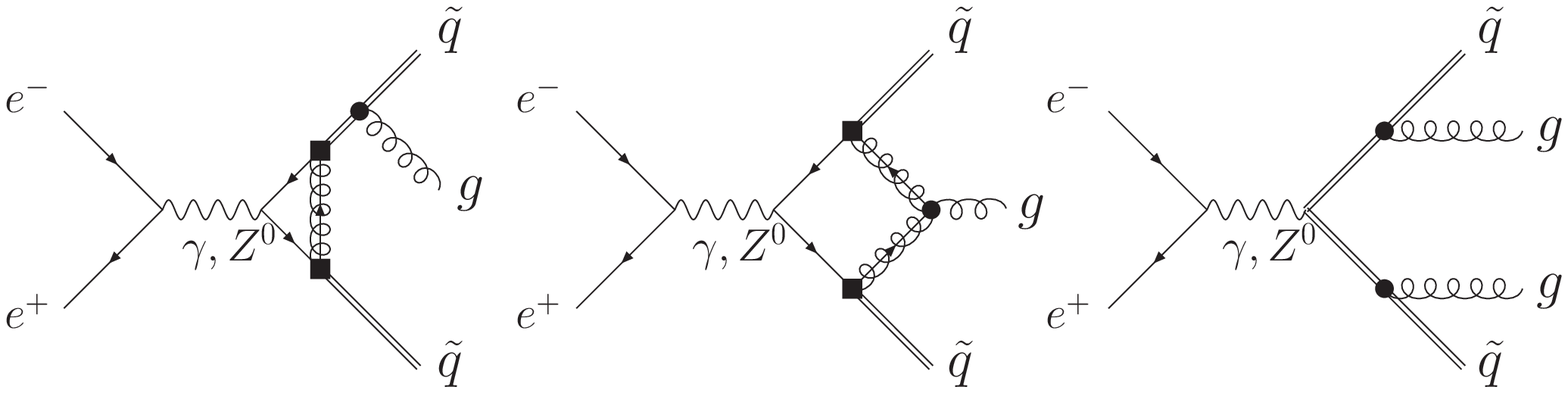}
\caption{\label{diag-sqsqgl} {\it
Example diagrams for SUSY-QCD corrections to
squark pair plus gluon production, left:
gluino vertex correction, mid: box diagram;
right: double gluon bremsstrahlung. }}
\end{figure}

\begin{figure}[h!t] 
\centering
\includegraphics[width=0.5\linewidth]{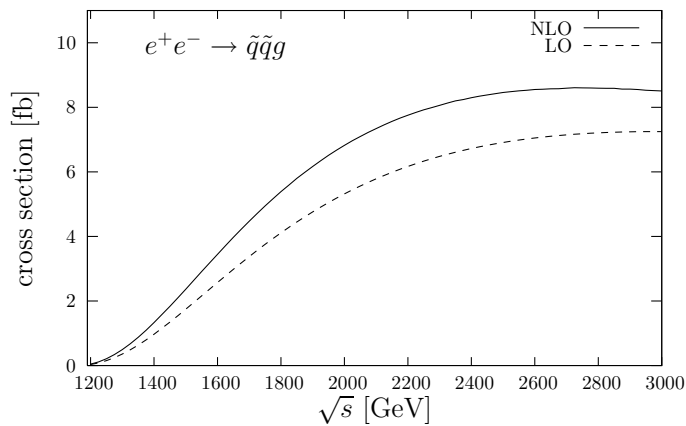}
\caption{\label{fig-sqsqgl-tot} {\it
Total cross section of the process $e^+e^- \to \qw \qw g$
for the c.m. energy $\sqrt{s}$ at leading
order [dashed line] and next-to-leading order [full line].
The mass parameters are adopted from SPS1a.
A minimal energy $E_X >$ 100 {\rm{GeV}} is assumed for single
jets $g$ and jet pairs $\,gg, \,\bar{q}q$ emitted off the squarks.}}
\end{figure}

In the numerical example, Fig.~\ref{fig-sqsqgl-tot}, we assume a minimal energy to be carried 
by the jet(s) emitted in addition to the squark pair.
The minimal energy  $E_X > 100$~GeV required for $X$ in the final state $\qw \qw X$
cuts out the infrared divergence in the gluon energy for
$X=g$ in Born approximation. The dashed line in Fig.~\ref{fig-sqsqgl-tot} corresponds to the Born
approximation. The solid line adds the complete set of SUSY-QCD radiative corrections.
These corrections include 
vertex and box diagrams as well as additional gluon-pair and light quark-antiquark
final states.
For these final states, $X=gg \; {\rm and} \; q{\bar{q}}$, 
the same cut-off in the energy ($E_X > 100$~GeV) is applied as before. 
Example diagrams for the radiative 
corrections are shown in Fig.~\ref{diag-sqsqgl}.
The semi-inclusive definition integrates out both 
the infrared and collinear singularities in the NLO corrected cross section. \\ 
In Born approximation the cross section depends strongly on the
renormalization scale introduced solely by the gauge coupling. 
The dependence is significantly reduced if the higher order corrections are included.
For example, varying the renormalization scale $\mu$ between $\frac{1}{5} \sqrt{s} \leq
\mu \leq 5 \sqrt{s}$  around a c.m. energy of $\sqrt{s}=3$~TeV, the cross section varies by $\Delta \sigma /
\sigma \simeq 14$\% in Born approximation, while the variation is damped to $4$\% in NLO
approximation. \\

In the peak region we find a sizeable cross section of several femtobarn. The 
radiative corrections enhance the cross section at a level of 20\%. 

%
\subsection[Squark-gluino final states $\mathbf{q \qw \gw}$]
{The golden channel: Squark-gluino final states $\mathbf{q \qw \gw}$}
%

\noindent
The golden channel for measuring the SUSY-QCD \mbox{$q \qw \gw$}
Yukawa coupling ${\hat{g}}_s$ {\it directly} is the process
\begin{equation}
         e^+e^- \to q \qw^\ast \gw + c.c.
\end{equation}
As evident from Fig.~\ref{LOdiagrams}, the size of the cross section is governed by the
quadratic dependence on the Yukawa coupling ${\hat{g}}^2 = 4 \pi {\hat{\alpha}}_s$:
\begin{equation}
\begin{split}
 \frac{d^2\sigma}{dx_q\:dx_{\qwb}} =&
   \frac{2\alpha^2\hat{\alpha}_s}{3s} C_{\LlR}
   \biggl\{
  \frac{1}{(1-x_{\qwb})^2} y_{\qwb}^2 (1-x_{\qwb}-\mu_{\gw}^2 + \mu_{\qwb}^2) \\
                                             &+ \frac{1}{(1-x_{q})^2} 
  \big[ 3(1-x_{q})(1-x_{\gw}) + 3\mu_{\gw}^2 - 3\mu_{\qwb}^2 - x_{q}^2(\mu_{\gw}^2 - \mu_{\qwb}^2)
   - \frac{1}{2}(1-x_{q})(y_q^2-y_{\gw}^2+y_{\qwb}^2) \big] \\ 
                                             &+ \frac{1}{(1-x_{q})(1-x_{\qwb})} \left[ x_{q} y_{\qwb}^2 
      - \frac{1}{2}(x_{\qwb}+2\mu_{\gw}^2 - 2\mu_{\qwb}^2)(y_q^2-y_{\gw}^2+y_{\qwb}^2) \right] \biggr\},
\end{split}
\end{equation}
in the standard notation introduced before. \\ 

The Born cross sections for both the cases $\mqw < \mgw$ and
$\mqw > \mgw$ are presented in Figs.~\ref{fig:LOcross}.
The processes give rise to $4$, respectively $6$ jets in the final state [at Born level],
with the 4-jet final state characteristic for $q \tilde{q} \tilde{g}$
production in the mass range $m_{\tilde{q}} < m_{\tilde{g}}$.\\
\begin{figure}[h!t] 
\centering
\vspace{5mm}
\includegraphics[width=0.65\linewidth,clip]{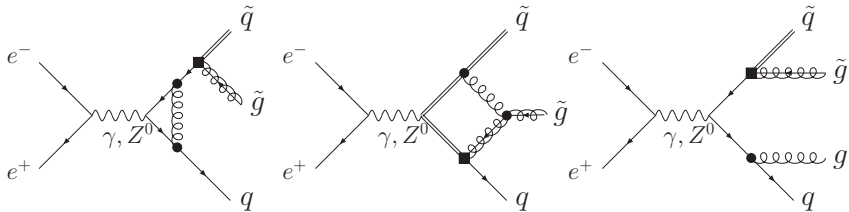}
\caption{\label{diag-qsqgli} {\it
Example diagrams for SUSY-QCD corrections to
squark gluino production, left:
gluon vertex correction, mid: box diagram;
right: real gluino emission. }}
\end{figure}
Examples  of NLO diagrams contributing to the SUSY-QCD corrections to $q \qw \gw$
final states are depicted in Fig.~\ref{diag-qsqgli}. Vertex and box
diagrams are built up by gluinos, gluons, squarks and quarks. To order $\alpha_s$, 
4-parton final states are generated only by the emission of an additional gluon.
Since the radiatively corrected cross section is defined inclusively, infrared 
and collinear singularities are integrated out.  \\

\begin{figure}[h]
\centering
\begin{minipage}[b]{0.49\linewidth}
\includegraphics[width=\linewidth]{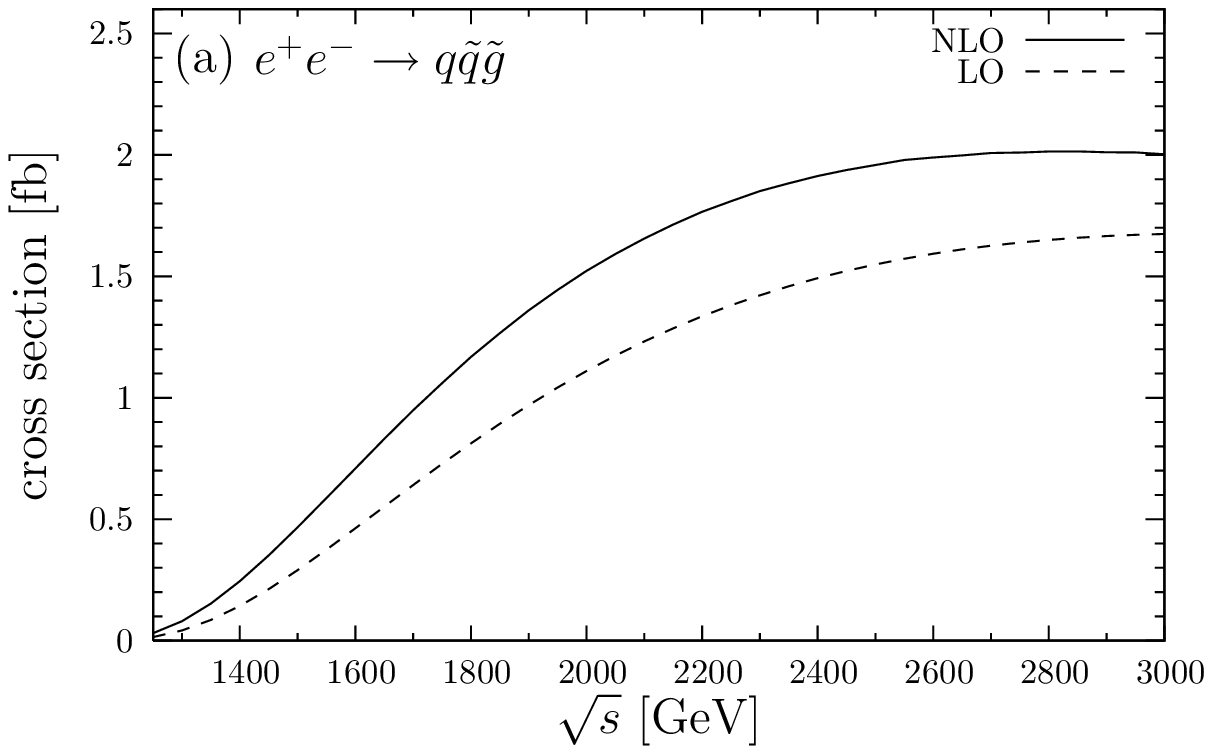}
\end{minipage}
\hfill
\begin{minipage}[b]{0.49\linewidth}
\includegraphics[width=\linewidth]{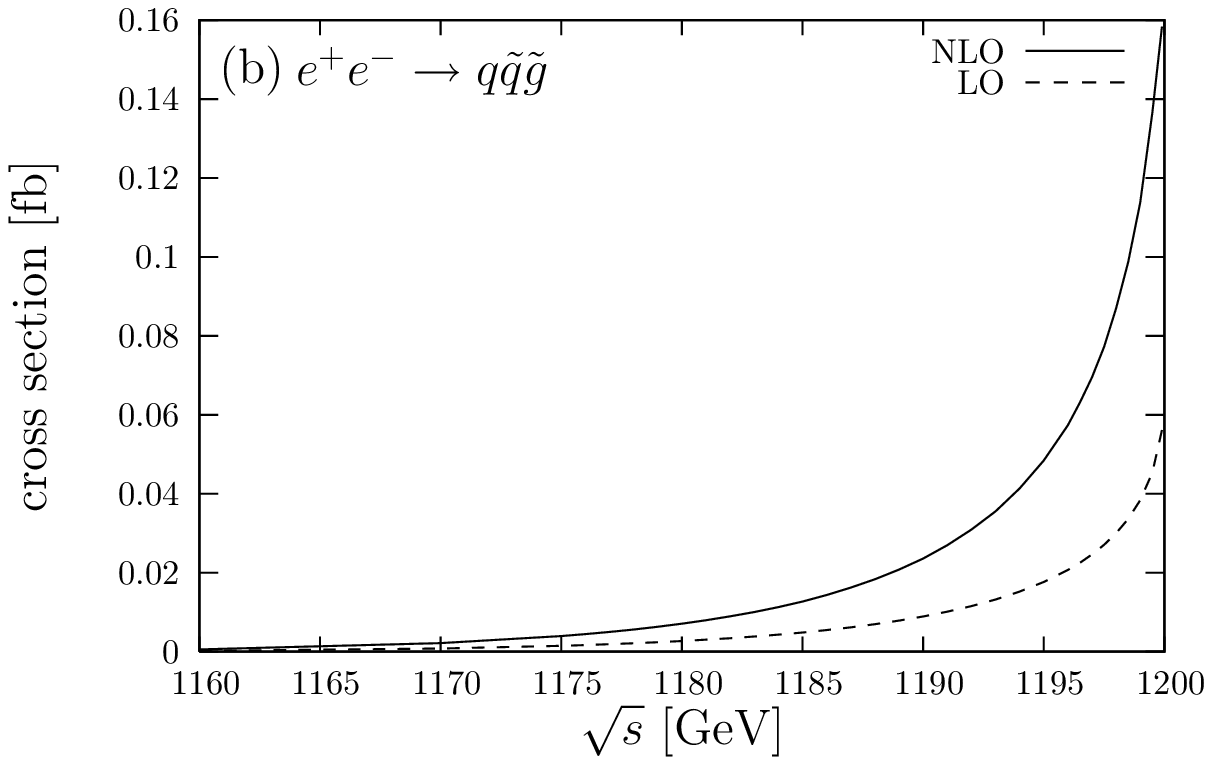}
\end{minipage}
\caption{\label{fig.qsqgli1} {\it (a) Total cross section for squark-gluino production 
at LO and NLO for SPS1a mass parameters, i.e.
$m_{\tilde{q}} < m_{\tilde{g}}$;\\ 
(b) for reversed SPS1a mass parameters, i.e. $m_{\tilde{q}} > m_{\tilde{g}}$.
[Above a c.m. energy
of twice the squark mass, squark-pair production
followed by squark decays to quarks and gluinos, is by far dominant.]}}
\end{figure}

The corrected cross sections are displayed in Figs.~\ref{fig.qsqgli1}
for squarks lighter and heavier than gluinos [SPS1a and 
SPS1a reversed], respectively. Again, the strong
dependence on the renormalization scale in Born approximation is damped
once the higher order corrections are included properly: 
If the renormalization scale $\mu$ varies between 
$\frac{1}{5} \sqrt{s} \leq
\mu \leq 5 \sqrt{s}$, for $\sqrt{s}=3$~TeV the cross section for $m_{\tilde{q}} < m_{\tilde{g}}$ 
varies by $\Delta \sigma / \sigma \simeq 14$\% in Born approximation, 
while the variation is damped to~$4$\% in NLO approximation. \\

The cross sections for $m_{\qw} < m_{\gw}$ are enhanced by the radiative 
corrections to a size of order $2$~fb which can be considered sufficient 
for experimental analysis if integrated luminosities $\sim 1$ ab$^{-1}$
can be achieved. In the mass range $m_{\qw} > m_{\gw}$ the size of the cross
section is rather small, however.\\

%
\subsection{Final states without supersymmetric particles}
%
\subsubsection{Quark pairs $\mathbf{qq}$}

\noindent
The basic quark-pair production process, defined in the absence of any
supersymmetric particle in the final state,
\begin{equation}
      e^+e^- \to q \bar{q}
\end{equation}
scales asymptotically with the energy squared:
\begin{equation}
\left.\sigma(e^+e^- \to q \bar{q})\right|_\text{Born}=  
\frac{4 \pi \alpha^2}{s}
\left[
	Q_q^2 + ({\cve}^2+{\cae}^2) ({\cvq}^2+{\caq}^2) \chi_Z^2 
    - 2 Q_q \cve \cvq \chi_Z
\right]
\,.
\end{equation}

\begin{figure}[h!t] 
\centering
\includegraphics[width=0.6\linewidth,clip]{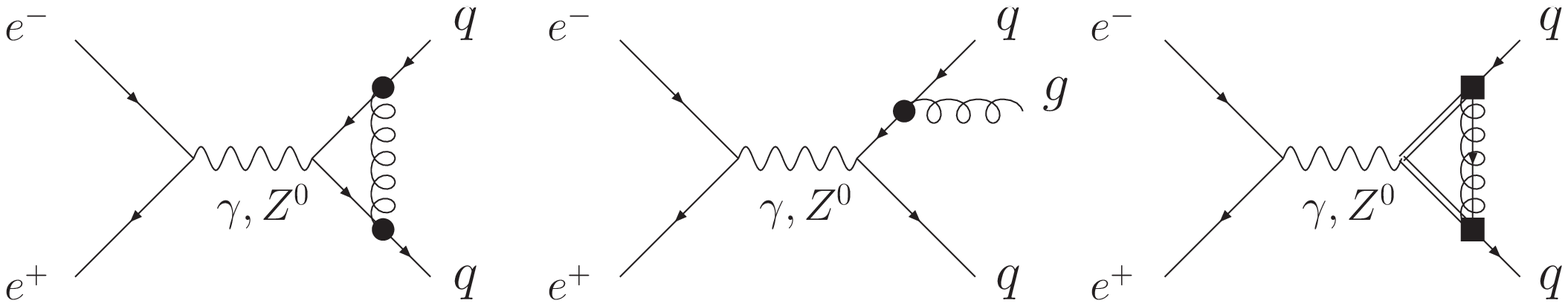}
\caption{\label{diag-qq} {\it
Examples for diagrams of QCD and SUSY-QCD corrections to
the production of quark pairs; left:
gluon vertex correction, mid: gluon bremsstrahlung;
right: squark/gluino vertex correction. }}
\end{figure}
\begin{figure}[h!]
\centering
\includegraphics[width=0.47\linewidth]{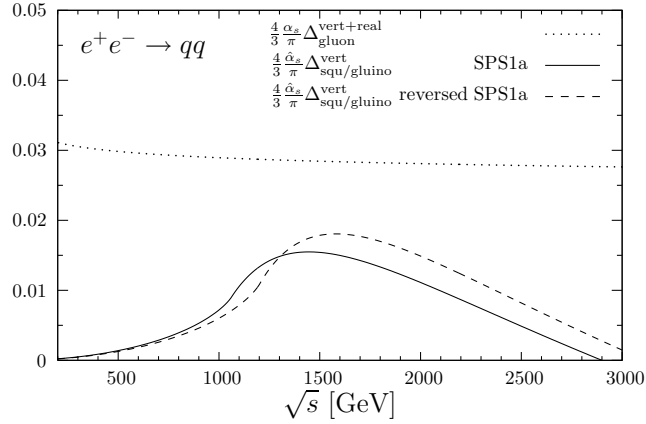}
\caption{\label{fig.qq} {\it NLO QCD, $(4 \alpha_s/ 3 \pi) \Delta^\text{vert+real}_\text{gluon}$, 
as well as NLO SUSY-QCD, $(4 {\hat{\alpha}}_s / 3 \pi) \Delta^\text{vert}_\text{squ/gluino}$, corrections
of the total cross section for quark pair production
in the reference points SPS1a and reversed SPS1a.}}
\end{figure}

The Born cross section is modified, to NLO, by gluon exchanges in the vertex 
and by gluon bremsstrahlung, as well as squark/gluino corrections to the vertex,
Fig.~\ref{diag-qq}, 
\begin{equation}
\sigma_\text{NLO}(e^+e^- \to q \bar{q})= 
\left.\sigma(e^+e^- \to  q \bar{q})\right|_\text{Born} 
\left(1 + \frac{4}{3}\frac{\alpha_s}{\pi} \Delta^\text{vert+real}_\text{gluon} 
      + \frac{4}{3}\frac{{\hat{\alpha}}_s}{\pi} \Delta^\text{vert}_\text{squ/gluino} \right) \, .
\end{equation}
The gluon correction to NLO is well known, $\Delta^\text{vert+real}_\text{gluon}= 3/4$, while 
the genuine triangular squark/gluino correction
is more involved \cite{Hagiwara:1990st}.
For the set of SPS1a and reversed SPS1a mass
parameters introduced before, the pure QCD corrections are in general leading,
except for very high energies.
The individual gluon and SUSY corrections
are presented in Fig.~\ref{fig.qq}. [The two contributions are
separately gauge invariant.] 
The Born cross section drops from $2.2$~pb down to $61$~fb in the range
of c.m.~energies between $0.5$ and $3$~TeV. 
The genuine SUSY-QCD corrections involving the Yukawa coupling of squarks to
quarks and gluinos are moderately large and of half the size of the standard QCD corrections. 
In the peak region the radiative SUSY corrections
are of the order of $1.5$\%; the corrections flip sign at an energy of
$\sim 3$~TeV for the two reference sets of SUSY masses. 
For $\mqw=\mgw \equiv \tilde{M}$ the correction 
can be cast in the compact form
\begin{equation}
\Delta^\text{vert}_\text{squ/gluino} =
\frac{3}{4}+ \frac{\beta}{4} \ln \left(\frac{1-\beta}{1+\beta}\right)
+\frac{1}{16} (1-\beta^2) \left [
\ln^2 \left(\frac{1-\beta}{1+\beta}\right) - \pi^2 \right]
\,,
\end{equation}
where $\beta = (1-4 \tilde{M}^2/s)^{1/2}$.
The SUSY correction approaches
\begin{equation}
\Delta^\text{vert}_\text{squ/gluino} =
 - {1}/{4}\, \ln(s/{\tilde{M}}^2)
\end{equation}
for asymptotic energies. The logarithmic behavior in the energy is consistent
with the Lee-Nauenberg theorem since gluino radiation is not included in the final
state; in fact, adding gluino radiation cancels the logarithm in the asymptotic region.

\subsubsection{Quark Pairs and Gluons $\mathbf{qqg}$}

\begin{figure}[h!] 
\centering
\vspace{5mm}
\includegraphics[width=0.65\linewidth,clip]{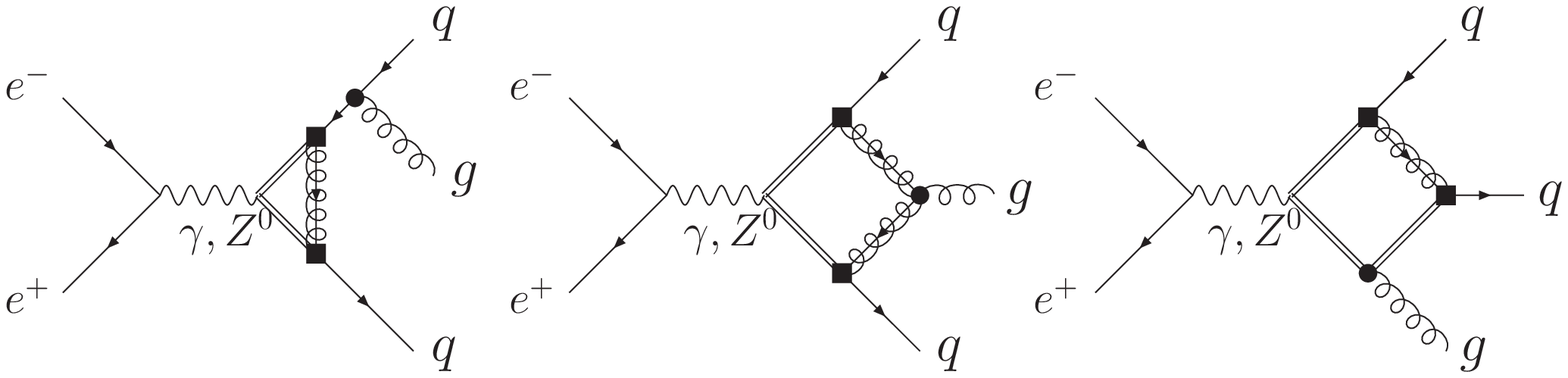}
\caption{\label{diag-qqg} {\it
Examples for diagrams of SUSY-QCD corrections to
the production of quark pairs accompanied by a gluon; left:
squark/gluino vertex correction, mid and right: squark/gluino box correction. }}
\end{figure}

\begin{figure}[h!]
\centering
\includegraphics[width=0.45\linewidth]{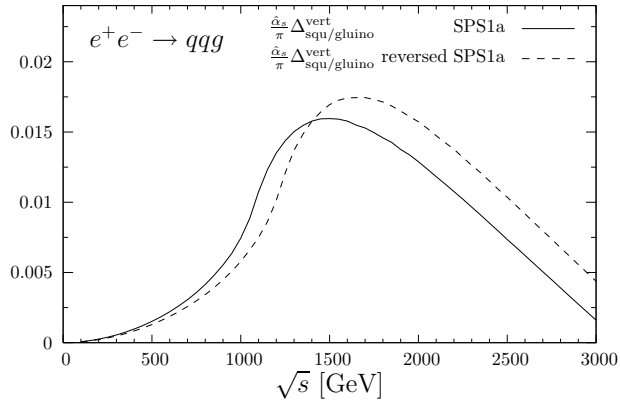}
\caption{\label{fig.qqg} {\it NLO SUSY-QCD corrections, 
$({\hat{\alpha}}_s/\pi)\Delta^\text{vert}_\text{squ/gluino}$, 
to the total cross section for $q \bar{q} g$ production for
the mass hierarchies of SPS1a, solid line, and SPS1a reversed, dashed line.}}
\end{figure}

\noindent
If additional gluon jets are isolated in the final state,
\begin{equation}
      e^+e^- \to q \bar{q} g \,,
\end{equation}
the analysis becomes much more complex. Starting from the Dalitz plot
density for gluon radiation \cite{r14},
\begin{equation}
\left.\frac{d^2 \sigma (e^+e^- \to q \bar{q} g)}{dx_q dx_{\bar{q}}}\right|_\text{Born} 
= 
\left.\sigma (e^+e^- \to q \bar{q})\right|_\text{Born} \;
\frac{2 \alpha_s}{3 \pi} 
\frac{x_q^2 + x_{\bar{q}}^2}{(1-x_q) (1-x_{\bar{q}})}\,,
\end{equation}
with the scaled energies $x_q = 2 E_q/\sqrt{s}$ etc,
the jet events are defined for a cut in the invariant masses of the parton pairs
$y_{ij} = (p_i + p_j)^2/ s > 0.05$.
The radiative corrections may be decomposed in the form 
\begin{equation}
\sigma_\text{NLO}(e^+e^- \to q \bar{q} g)= 
\left.\sigma(e^+e^- \to \bar{q} q g)\right|_\text{Born} 
\left(1 + \frac{\alpha_s}{\pi}\Delta^\text{vert+real}_\text{gluon} 
+ \frac{{\hat{\alpha}}_s}{\pi}\Delta^\text{vert}_\text{squ/gluino}\right) \,.
\end{equation}
As generally anticipated for corrections to jet cross-sections, 
the gluon corrections, {\it c.f.} Ref.~\cite{B2},
\begin{equation}
\frac{\alpha_s}{\pi}\Delta^\text{vert+real}_\text{gluon} \simeq
     \frac{\alpha_s}{\pi} \frac{B}{2A}
\end{equation}
are by far dominant with $B/2A \simeq$ 10 for $y_{ij}$ = 0.05, {\it i.e.}
$\frac{\alpha_s}{\pi}\Delta^\text{vert+real}_\text{gluon} \simeq 0.3$.
The genuine SUSY-QCD corrections 
are defined as any corrections involving propagators of supersymmetric
particles, some examples are shown in Fig.~\ref{diag-qqg}.
Such diagrams come with two powers of the quark-squark-gluino
Yukawa coupling at next-to-leading order. The genuine SUSY contribution 
$\frac{{\hat{\alpha}}_s}{\pi}\Delta^\text{vert}_\text{squ/gluino}$
is displayed
in Fig.~\ref{fig.qqg}. 
The two mass hierarchies $\mqw < \mgw$ and $\mqw > \mgw$ are adopted
again from the reference point SPS1a and the reversed SPS1a point.
In the respective peak regions SUSY corrections of about~$1.5$\%
are predicted, significantly smaller than the standard QCD corrections,
as anticipated.

%
\section{Assessment}
%
%
%
%
\begin{table*}[t]
\centering
\[\begin{array}{cccccccc}
\hline\hline
   &  \text{Born coupling} 
   & {\sqrt{s}}\{\text{max}\}{\rm [GeV]} 
   & \sigma_{\text{Born}}\, {\rm [fb]}
   & \sigma_{\text{NLO}}\, {\rm [fb]}
   & \frac{4}{3} \frac{{\alpha}_s}{\pi} \Delta_{\text{QCD}}
   & \frac{4}{3} \frac{\hat{\alpha}_s}{\pi} \Delta_{\text{SUSY}}
   & {\Delta \hat{\alpha}_s}/{\hat{\alpha}_s}\\
\hline
\qw \qw   &                & 3,000 & 25.1 & 26.0 & 0.126 & -0.09  & 7.0\%\\
q q       &                & 1,300 & 328  & 342	& 0.0277 & 0.0159 & 11\%\\
q \qw \gw & \hat{\alpha}_s & 2,850 & 1.66 & 2.01& 0.661	& -0.448  & 2.5\%\\   
\hline
\hline
   & 
   & 
   & 
   &
   & \frac{{\alpha}_s}{\pi} \Delta_{\text{QCD}}
   & \frac{\hat{\alpha}_s}{\pi} \Delta_{\text{SUSY}}
   & {\Delta {\alpha}_s}/{{\alpha}_s}\\
\hline
\qw \qw g & \alpha_s   & 2,800 & 7.21 & 8.59 & 0.276 & -0.0836 & 1.2\%\\
q q g     & \alpha_s   & 2,800 & 10.5 &	13.1 & 0.257 & 0.0039 & 1.0\%\\
\hline\hline
\end{array}\]
\caption{\label{tab:sensalphas}
  {\it 1-sigma statistical errors of the Yukawa $\hat{\alpha}_s$  and gauge $\alpha_s$ 
  couplings determined from quark-squark-gluino final states and SUSY radiative corrections  
  in various channels based on SPS1a SUSY parameters. The upper part of the table with 
  the final states $\qw \qw$, $q q$ and $q \qw \gw$ describes the measurement of $\hat{\alpha}_s$,
  and the lower part, with $\qw \qw g$ and $qqg$, the measurement of $\alpha_s$. 
  The c.m. energy $\sqrt{s}\{\text{max}\}$ is chosen, below 3 TeV, 
  such that the sensitivity to the measurement of the couplings is maximized,
  except for the gauge coupling in $qqg$ where the scale is adjusted to $\qw \qw g$.
  }}
\end{table*}
\noindent
In the preceding section the theoretical base has been studied for
future measurements of the Yukawa coupling between quarks, squarks and
gluinos in high-energy $e^+ e^-$ collisions. Supersymmetry predicts 
this coupling to be identical with the standard QCD gauge coupling between 
quarks and gluons, and squarks and gluons. Various channels have been
investigated which can be explored to test this fundamental identity
between the couplings, either indirectly through virtual corrections
to quark and squark pair production in $e^+ e^-$ collisions, or directly
in comparing the cross section for the golden channel $q \qw \gw$ with
the standard gluon radiation processes $qqg$ and $\qw \qw g$. While the 
golden channel measures the $q \qw \gw$ Yukawa coupling ${\hat{g}}_s$, the 
radiation processes $\qw\qw g$ and $qqg$ determine the QCD gauge coupling $g_s$ 
in the squark sector and the standard quark sector for comparison.\\

\psfrag{msquark}{$\mqw$ [GeV]}
\psfrag{mgluino}{$\mgw$ [GeV]}
\psfrag{deltaalphas}{\hspace{4mm}\large ${\Delta \hat{\alpha}_s}/{\hat{\alpha}_s}$}
\begin{figure}[h!] 
\centering
\begin{minipage}[b]{0.3\linewidth}
  $e^+e^- \to q q$, $\sqrt{s}<1$~TeV
\end{minipage}
\hfill
\begin{minipage}[b]{0.3\linewidth}
  $e^+e^- \to q \qw \gw$, $\sqrt{s}<1$~TeV
\end{minipage}
\hfill
\begin{minipage}[b]{0.3\linewidth}
  $e^+e^- \to q \qw \gw$, $\sqrt{s}<3$~TeV
\end{minipage}
\\
\begin{minipage}[b]{0.3\linewidth}
\includegraphics[width=\linewidth]{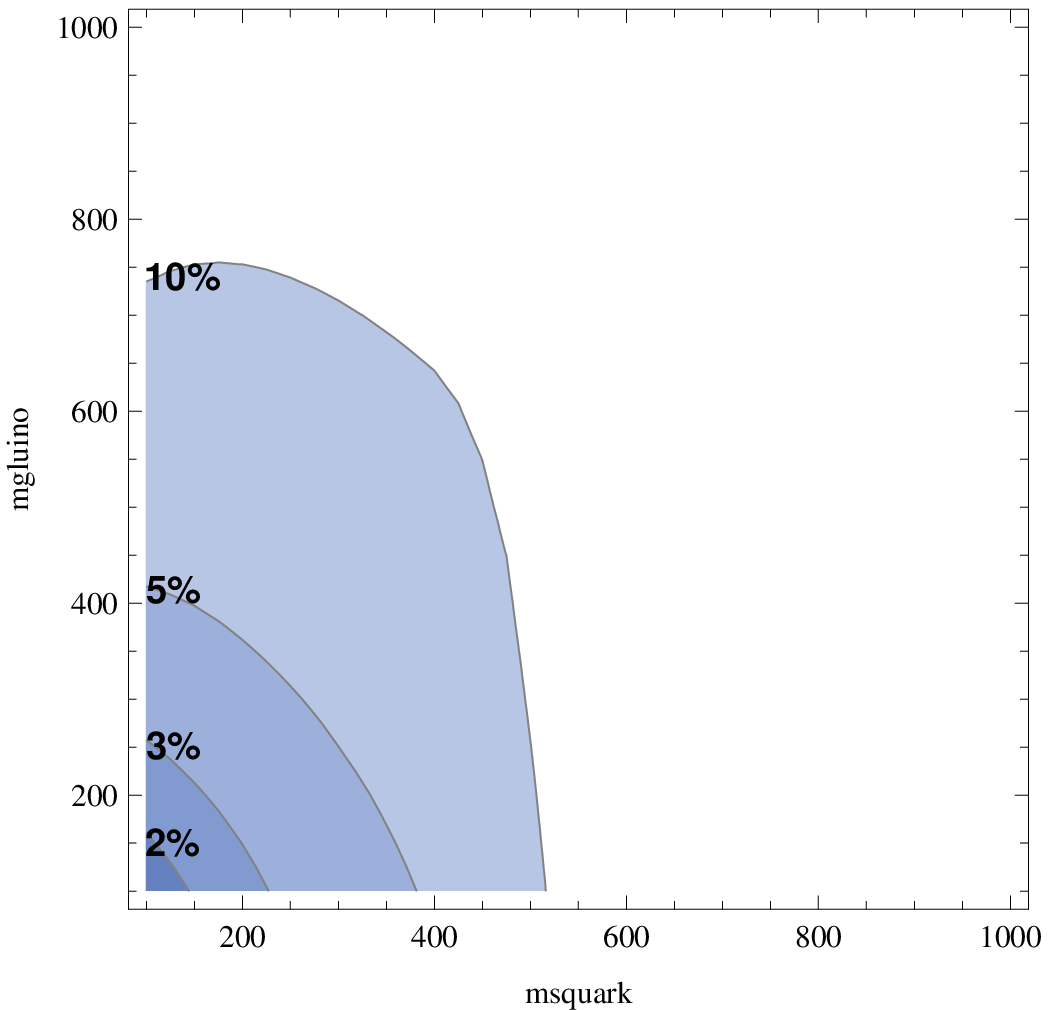}
\end{minipage}
\hfill
\begin{minipage}[b]{0.3\linewidth}
\includegraphics[width=\linewidth]{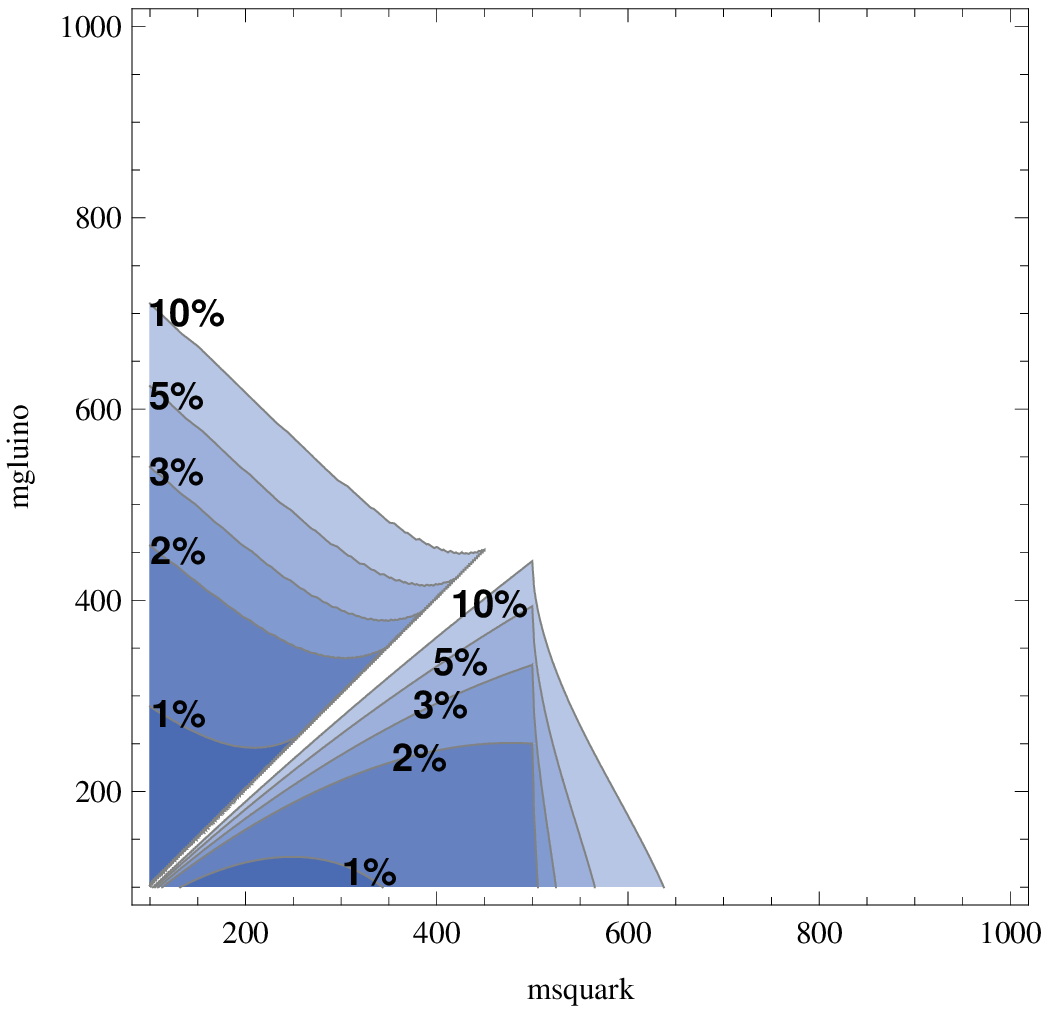}
\end{minipage}
\hfill
\begin{minipage}[b]{0.3\linewidth}
\includegraphics[width=\linewidth]{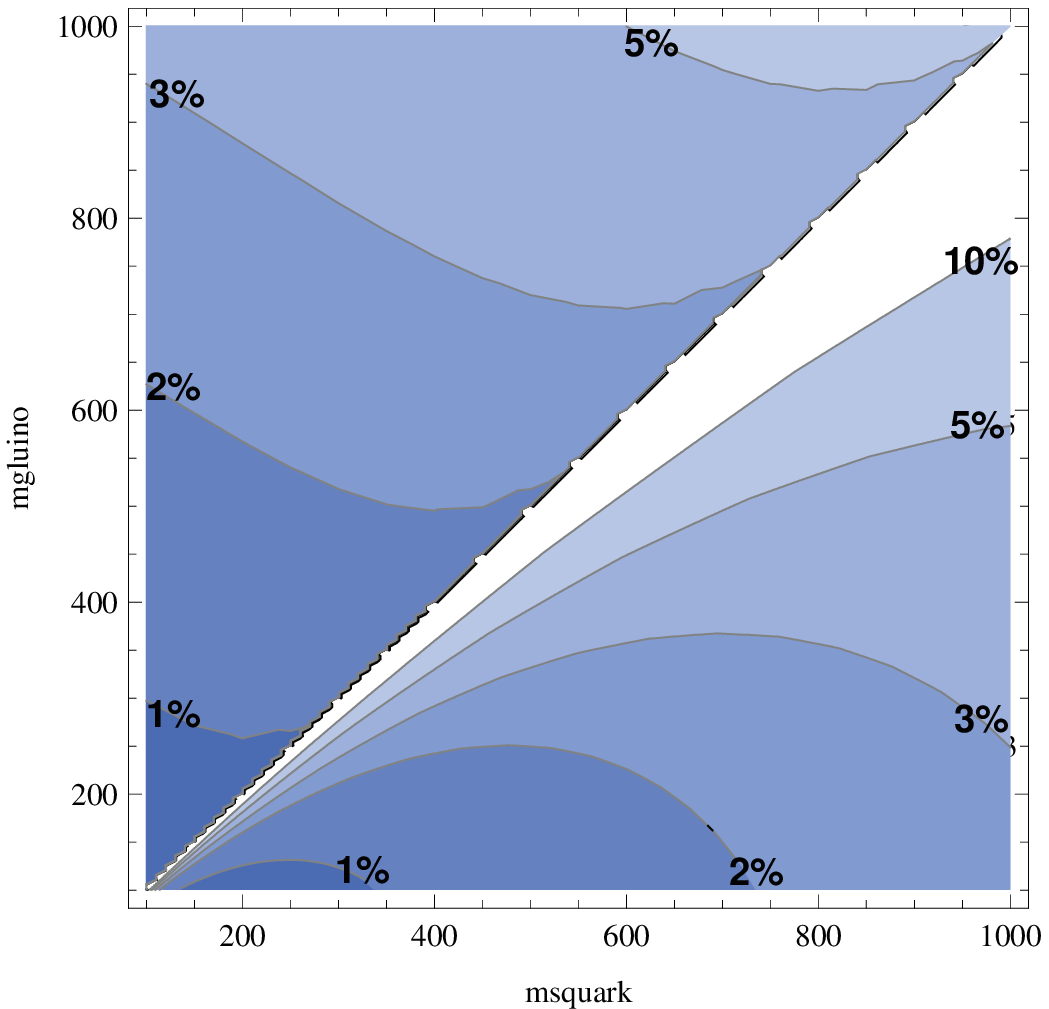}
\end{minipage}
\caption{\label{fig.contalphas-gold} {\it
Contours of the statistical 
1-sigma errors of the Yukawa coupling ${\Delta \hat{\alpha}_s}/{\hat{\alpha}_s}$ 
in the $[\mqw,\mgw]$ mass plane, {\rm Left}: for the indirect channel
$e^+e^- \to q q$ at ILC energies; {\rm Mid, Right}: for the direct golden channel
$e^+e^- \to q \qw \gw$, corresponding to a maximal ILC c.m. energy of $1$~{\rm TeV}, 
and to $3$~{\rm TeV} for CLIC, respectively. The blind wedge can be covered by
analyzing the partial widths. }}
\end{figure}
The central results are summarized in Tab.~\ref{tab:sensalphas}
and Figs.~\ref{fig.contalphas-gold}. 
In the table the SUSY parameters are adopted from the reference point SPS1a 
with squark and gluino masses of 530 GeV and 600 GeV, respectively. 
Since by the time of measurements of the couplings the squark and gluino masses
will be determined accurately,
special values of the $e^+ e^-$ c.m. energy
can be chosen such as to maximize the sensitivity for the 
measurements of the Yukawa and gauge couplings.
A standard value $\int\!{\mathcal{L}} 
= 1$~ab$^{-1}$ is taken for the integrated luminosity. The errors quoted 
in the table are purely {\it statistical}, reflecting the minimal errors 
that can be expected in any future experimental analysis.\\

The left-most figure in the panel Figs.~\ref{fig.contalphas-gold}
displays the contours of the {\it statistical} 1-sigma errors
${\Delta \hat{\alpha}_s}/{\hat{\alpha}_s}$ for the standard $qq$
channel in the $[\mqw,\mgw ]$ mass plane for ILC, irrespective
of any specific model except SUSY-QCD. For each $\qw,\gw$ mass pair  
the collider energy is chosen such as to provide maximal sensitivity
to the measurement of ${\hat{\alpha}}_s$. The Yukawa coupling 
${\hat{\alpha}}_s$ can be determined in this channel only indirectly
through virtual SUSY corrections. Such analyses
may provide a first glimpse of the Yukawa coupling. \\

\noindent
The figures in the middle and on the right of the panel summarize {\it statistical} 
analyses of the golden $q \qw \gw$ channel on the same basis. In this channel 
the Yukawa coupling can be measured directly since the
Born cross section is  proportional to ${\hat{\alpha}}_s$.  
The results are presented for ILC c.m. energies up to a maximum of 1 TeV,
and for CLIC up to 3 TeV. As before, the energies are chosen at the
point of maximal sensitivity for the Yukawa coupling for any mass
pair $[\mqw,\mgw]$. For $\mgw < \mqw$ the collider energy is
restricted to the range below the $\qw\qw$ threshold as to avoid
contamination of the $q\qw\gw$ channel due to on-shell $\qw \to q \gw$ decays.
This restriction generates a blind wedge near the threshold which however
can be covered in analyses of the partial widths as discussed before. \\

The estimate of the statistical error $\Delta {\hat{g}}_s / {\hat{g}}_s \simeq 1\%$
for the measurement of the Yukawa coupling in this report is only a first step. 
The step is promising, nevertheless, as the expected error is sufficiently small. 
Even if an order of magnitude of events is lost due to experimental 
cuts and efficiencies in controlling, for example, the chargino and neutralino 
modes of squark decays [and cascading gluinos], the
resulting final error is still at a level of 5 to 10\%. This size
compares reasonably well with expectations for LHC experimental analyses
in Ref.~\cite{Freitas:2007fd} in which squark/gluino decays are included explicitly.
Thus, LHC and a TeV lepton collider are complementary instruments for the involved 
measurements of the SUSY-QCD Yukawa and gauge couplings. In particular, since 
the relevant cross sections drop with rising squark and gluino masses quadratically
and faster{\footnote{Since analytical expressions for cross sections 
at next-to-leading order are quite involved, they are not noted here explicitly, 
but they can be retrieved from 
\textsf{http://www.thphys.uni-heidelberg.de/$\sim$maniatis/susyqcd}.}}, 
mutual complementarity 
of the two methods raises prospects of testing the identity of Yukawa and 
gauge couplings in super-QCD considerably. \\

%
\acknowledgments{\noindent Thanks go to M.~Spira, S.~Dittmaier and D.~Zerwas for 
                 helpful discussions. PMZ is grateful to the Inst.~Theor.~Phys.~E 
                 for the warm hospitality extended to him at RWTH Aachen.
		 The work of AB was supported by a Heisenberg grant of the DFG.
                 The work of MW was supported in part by the National Science Foundation
		 under grant NSF-PHY-0547564.}

%
%
%

\end{document}